\documentclass[twoside,a4paper,11pt]{sca}
\def\apj{ApJ}%
\def\apjl{ApJL}%
\def\apjs{ApJS}%
\def\mnras{MNRAS}%
\usepackage{graphicx,amssymb}
\usepackage{hyperref}
\usepackage{movie15}
\usepackage{natbib}  
\begin{document}
\pagenumbering{arabic}
\pagestyle{myheadings}
\thispagestyle{empty}
{\flushright\includegraphics[width=\textwidth,bb=90 650 520 700]{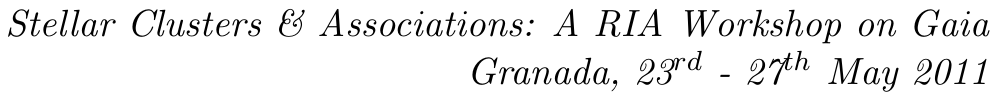}}
\vspace*{0.2cm}
\begin{flushleft}
{\bf {\LARGE
%
Probing the Microscopic with the Macroscopic: from Properties of Star Cluster Systems \\ to Properties of Cluster-Forming Regions

%
}
\vspace*{1cm}
%
Genevi\`eve Parmentier$^{1,2}$
%
}\\
\vspace*{0.5cm}
%
$^{1}$
Max-Planck-Institut f\"ur Radioastronomie, Bonn, Germany \\
$^{2}$
Argelander-Institut f\"ur Astronomie, Bonn, Germany \\
%
\end{flushleft}
%
\markboth{
Properties of Cluster-Forming Regions
}{ 
%
Genevi\`eve Parmentier
%
}
\thispagestyle{empty}
\vspace*{0.4cm}
\begin{minipage}[l]{0.09\textwidth}
\ 
\end{minipage}
\begin{minipage}[r]{0.9\textwidth}
\vspace{1cm}
\section*{Abstract}{\small
%

To understand how systems of star clusters have reached their presently observed properties constitutes a powerful probe into the physics of cluster formation, without needing to resort to high spatial resolution observations of individual cluster-forming regions (CFRg) in distant galaxies. 
In this contribution I focus on the mass-radius relation of CFRgs, how it can be uncovered by studying the gas expulsion phase of forming star clusters, and what the implications are.  I demonstrate that, through the tidal field impact upon exposed star clusters, the CFRg mass-radius relation rules cluster infant weight-loss in dependence of cluster mass.  The observational constraint of a time-invariant slope for the power-law young cluster mass function is robustly satisfied by CFRgs with a constant mean volume density.  In contrast, a constant mean surface density would be conducive to the preferential destruction of high-mass clusters.  A purely dynamical line-of-reasoning leads therefore to a conclusion consistent with star formation a process driven by a volume density threshold.  Developing this concept futher, properties of molecular clumps and CFRgs naturally get dissociated.  This allows to understand: (i) why the star cluster mass function is steeper than the molecular cloud/clump mass function; (ii) the presence of a massive star formation limit in the mass-size space of molecular structures.

%
\normalsize}
\end{minipage}
%
%
%
\section{Introduction \label{sec:intro}}

We are still a long way from observationally mapping active or future cluster-forming regions on a one-by-one basis in distant galaxies.  Even at ALMA resolution (0.1 arcsec), to map clumps of cold dense molecular gas on a pc-scale is attainable only in galaxies less distant than $\simeq 2$\,Mpc.  In contrast, properties of {\it systems} of star clusters, such as the distribution functions of cluster ages, masses and radii, are retrievable out to distances of at least 20\,Mpc.  These distribution functions are shaped by both the dynamical evolution and the formation conditions of star clusters.  Therefore, evolving model cluster systems for a wide range of initial conditions, and comparing their predicted distribution functions to their  observed counterparts, constitutes a powerful tool to probe into the physics of cluster formation.  That `macroscopic' approach has the tremendous advantage to alleviate the need for a spatial resolution which is yet to be achieved in galaxies more distant than a few Mpc (`microscopic' approach).

Systems of young clusters are -- obviously -- those best suited to uncover cluster-forming region properties.  At ages less than 100\,Myr, a key-driver of cluster evolution is violent relaxation -- namely the dynamical response of embedded clusters to the expulsion of their residual star-forming gas due to massive star activity \citep[e.g.][]{2001MNRAS.323..988G}.  During violent relaxation, clusters expand and lose stars (infant weight-loss) as a result of the binding gaseous mass loss.  Initial conditions of this phase leave therefore signatures in the cluster age, mass and radius distribution functions \citep[e.g.][]{2008ApJ...678..347P, 2009ApJ...690.1112P}.  In this contribution, I highlight how the time-invariant shape of the {\it young} cluster mass function sheds light on the mass-radius relation of cluster-forming regions (hereafter CFRg).  I show that this relation is one of constant mean volume density, that is, $r_{CFRg} \propto m_{CFRg}^{1/3}$ with $r_{CFRg}$ and $m_{CFRg}$ the CFRg radius and mass, respectively.  Furthering this argument, I demonstrate: (i) why the young cluster mass function is steeper than the molecular cloud mass function, and (ii) why there is a massive star formation limit in the mass-size space of molecular structures.

\section{CFRg Mass-Radius Relation: a Dynamical Perspective {\label{sec:mrr}}}
\subsection{Tidal field impact implies constant mean volume density for CFRgs \label{ssec:tf}}

The amplitude of the cluster mass function is observed to steadily decrease with time over the first 100\,Myr of cluster evolution due to infant weight-loss.  Yet, its shape is reported to remain remarkably unchanged \citep[e.g.][]{2010ApJ...711.1263C}.  That is, cluster infant weight-loss is mass-independent.  When plotted as the number of clusters per linear mass interval, the cluster mass function is a featureless power-law of slope $-2$, ${\rm d}N \propto m_{\star}^{-2} {\rm d}m_{\star}$, irrespective of the cluster age bin (say, 1-10 or 10-100\,Myr).   

The mass fraction of stars remaining bound to a cluster by the end of its violent relaxation, $F_{bound}$, depends on the star formation efficiency of the parent CFRg, $SFE$, and on the gas expulsion time-scale, $\tau_{GExp}/\tau_{cross}$, expressed in units of the CFRg crossing-time: $F_{bound}=F_{bound}(SFE,\tau_{GExp}/\tau_{cross})$.  That infant weight-loss, $1-F_{bound}$, is mass-independent straightforwardly implies mass-independent $SFE$ \citep[fig.~1 in][see also Section 3.1]{2007MNRAS.377..352P}, and mass-independent $\tau_{GExp}/\tau_{cross}$ \citep{2008ApJ...678..347P}, although the constraint on $\tau_{GExp}/\tau_{cross}$ is looser than for the $SFE$ \citep[see section 4.1 in][for a discussion]{2011MNRAS.411.1258P}.  

A third, and so far overlooked, aspect is how the combination of an external tidal field with the mass-radius relation of CFRgs influences cluster infant weight-loss in dependence of cluster mass.  Due to gas-expulsion-driven expansion, cluster stars -- which, with no external tidal field, would remain bound -- may be driven beyond the cluster tidal radius and turned into field stars.  Therefore, infant weight-loss is partly governed by how deeply an embedded cluster sits within its limiting tidal radius, an effect quantified by the pre-expansion ratio of the cluster half-mass radius to tidal radius, $r_h/r_t$.  Accordingly, $F_{bound}=F_{bound}(SFE,\tau_{GExp}/\tau_{cross}, r_h/r_t)$.  The embedded-cluster half-mass radius scales with the CFRg radius, i.e. $r_h \propto \kappa.r_{CFRg}$, with the factor $\kappa$ depending on the assumed density profile.  The tidal radius, $r_t$, depends on both the embedded-cluster mass, $m_{ecl}$, and the strength of the external tidal field.  Equation~4 in \citet{2011MNRAS.411.1258P} gives $r_t$ for a cluster at a galactocentric distance $D_{gal}$ in an  isothermal galactic halo of circular velocity $V_c$:
\begin{equation}
\label{eq:rt}
r_t=\left(\frac{ G.D_{gal}^2}{2 V_c^2}.m_{ecl}\right)^{1/3}\;.
\end{equation}

\citet{2007MNRAS.380.1589B} build on the $r_h/r_t$ ratio to quantify how much an external tidal field enhances infant weight-loss compared to a tidal-field-free environment.  I will refer $r_h/r_t$ as the tidal field {\it impact}, to distinguish it from the tidal field strength.  The tidal field strength depends solely on the external tidal field, say, $V_c$ and $D_{gal}$.  These parameters also define the galactic halo volume density at a galactocentric distance $D_{gal}$, $\rho_{gal}(D_{gal})=V_c^2/(4\pi G D_{gal}^2)$, that is, in our example, the density of the CFRg environment.  In contrast, the tidal field impact depends on {\it both} the tidal field ($V_c$, $D_{gal}$) {\it and} the embedded-cluster properties ($m_{ecl}$, $r_h$), where the stellar mass of the embedded cluster obeys $m_{ecl}=SFE \times m_{CFRg}$.  In other words, the tidal field strength alone does not define the sensitivity of a cluster to the tidal field.  This is the {\it contrast  between the volume densities} of the CFRg and its environment which matters.  For an isothermal galactic halo, using eqs.~4 and 7 in \citet{2011MNRAS.411.1258P}, one can show that:
\begin{equation}
\label{eq:dct}
\frac{r_h}{r_t} = \kappa \left(\frac{2}{SFE}\right)^{1/3} \left(\frac{<\rho_{gal}(\leq D_{gal})>}{\rho_{CFRg}}\right)^{1/3}\,,
\end{equation}
where $<\rho_{gal}(\leq D_{gal})>$ is the mean volume density of the galactic halo {\it within} $D_{gal}$. 

When $r_h/r_t \lesssim 0.03$, the embedded cluster sits deeply within its limiting tidal radius, thereby giving rise to an expanded cluster `shielded' against the external tidal field. In other words, $F_{bound}(SFE,\tau_{GExp}/\tau_{cross}, r_h/r_t) \simeq F_{bound}(SFE,\tau_{GExp}/\tau_{cross})$: stars venturing beyond the tidal radius are lost because of velocities higher than the cluster escape velocity in the first place.  In contrast, when $r_h/r_t \gtrsim 0.20$, clusters evade tidal disruption only if their spatial expansion is damped severely through either high $SFE$ and/or long $\tau_{GExp}/\tau_{cross}$.  

Figure~3 of \citet{2011MNRAS.411.1258P} shows $F_{bound}$ in dependence of $m_{CFRg}$\footnote{Note that what we refer as `cluster-forming region' is called `cluster-forming core' in \citet{2011MNRAS.411.1258P}} for different CFRg mass-radius relations.  For the sake of simplicity, the assumed environment is an isothermal galactic halo.  It should be noted, however, that CFRgs are embedded in molecular clumps, themselves embedded in Giant Molecular Clouds (GMC).  The volume densities of these structures are higher than that of a galactic halo.  For instance, the mean volume density of a $\simeq 10^5\,{\rm M}_{\odot}$ GMC is $\simeq 1\,{\rm M}_{\odot}.{\rm pc^{-3}}$ \citep[assuming spherical symmetry and $\Sigma_{GMC} \simeq 40\,{\rm M}_{\odot}.pc^{-2}$,][]{2009ApJ...699.1092H}.  This is about 2 orders of magnitude denser than the density of an isothermal halo at $D_{gal} \simeq 8$\,kpc.  Therefore, $r_h/r_t$ and $F_{bound}$ values estimated by \citet{2011MNRAS.411.1258P} are lower and upper limits, respectively.  

\begin{figure}[!t]
\center
\includegraphics[width=15cm,angle=0,clip=true]{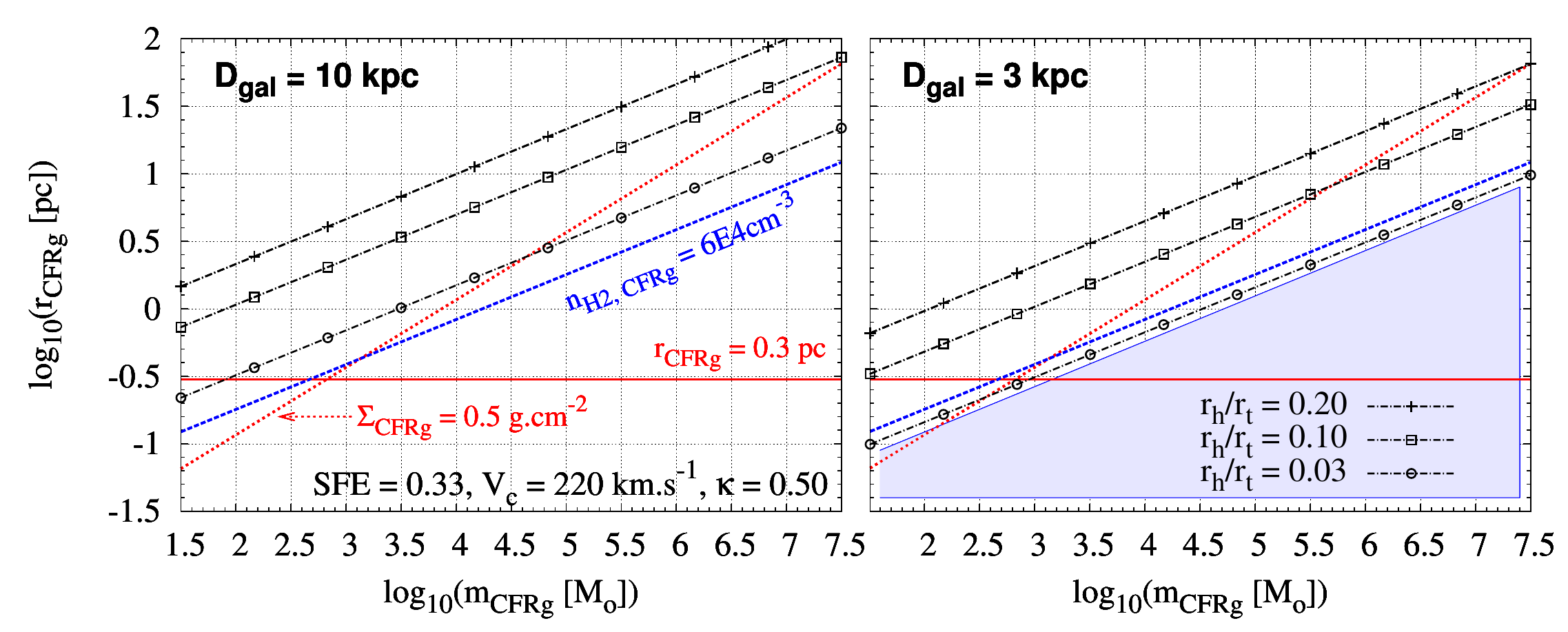} 
\caption{\label{fig:tf} 
Mass-radius diagrams of cluster-forming regions (CFRg) as a tool to assess the tidal field impact $r_h/r_t$  upon  clusters expanding after residual star-forming gas expulsion.  The parallelism between iso-$r_h/r_t$ lines (black lines with symbols; see legend) and lines of given mean number density $n_{{\rm H_2}, CFRg}$ (e.g. $n_{H_2, CFRg} = 6\times10^4\,cm^{-3}$, blue dashed line) demonstrates that $r_{CFRg} \propto m_{CFRg}^{1/3}$ is the most robust mass-radius relation to obtain mass-independent tidal field impact hence an invariant shape of the cluster mass function through violent relaxation.  In contrast, constant mean surface density and constant radius (e.g. $\Sigma_{CFRg} = 0.5\,g.cm^{-2}$ and $r_{CFRg} = 0.3\,pc$, red lines) lead to increasing and decreasing, respectively, tidal field impact with increasing CFRg mass.  When $r_h/r_t \lesssim 0.03$ (shaded area in right panel), cluster evolution is not or weakly only affected by the external tidal field.  Note that the iso-$r_h/r_t$ lines' intercepts depend on the tidal field strength (here the galactocentric distance, $D_{gal}$).      
}
\end{figure}

CFRgs of constant mean surface density ($r_{CFRg} \propto m_{CFRg}^{1/2}$) and constant radius ($r_{CFRg}$) are found to potentially lead to the preferential removal of high- and low-mass clusters, respectively.  In contrast, CFRgs of constant mean volume density ($r_{CFRg} \propto m_{CFRg}^{1/3}$) satisfies the observational constraint of mass-independent cluster infant weight-loss.  The physics driving these conclusions is best understood with a mass-radius diagram.  Figure~\ref{fig:tf} shows 3 CFRg mass-radius relations: one of constant radius $r_{CFRg}=0.3$\,pc (red solid line), one of constant mean surface density $\Sigma_{CFRg}=0.5\,g.cm^{-2}$ (red dotted line), and one of constant mean number density $n_{{\rm H_2}, CFRg}=6.10^4\,cm^{-3}$ (blue dashed line).  It also depicts 3 iso-$r_h/r_t$ lines in the $(m_{CFRg},r_{CFRg})$ space ($r_h/r_t=0.03, 0.10$ and $0.20$, black dash-dotted lines with symbols).  Both panels of Fig.~\ref{fig:tf} differ by the tidal field strength (here the galactocentric distance).  As Eq.~\ref{eq:dct} demonstrates, for a given environment, a given $r_h/r_t$ value equates with a line of constant mean volume density \citep[i.e. $r_{CFRg} \propto m_{CFRg}^{1/3}$, see also eq.~10 in][]{2011MNRAS.411.1258P}.  Note that iso-$r_h/r_t$ lines shift downwards when the tidal field gets stronger, that is, higher $\rho_{CFRg}$ are required to reproduce the same tidal field impact $r_h/r_t$. 

For all CFRgs lying below the line $r_h/r_t = 0.03$ (shaded area in the right panel), the tidal field impact is negligible.  In fact, for given $SFE$ and $\tau_{GExp}/\tau_{cross}$, $F_{bound}$ is mass-independent in that region of the plot, regardless of the mass-radius relation.  This is because $r_h/r_t=0.01$ or, say, $r_h/r_t=0.001$ affects $F_{bound}$ in the same way.  If CFRgs have masses and radii above the line $r_h/r_t=0.20$, cluster survival requires stringent conditions (high SFE, long $\tau_{GExp}/\tau_{cross}$).  In the intermediate region bound by $r_h/r_t=0.03$ and $r_h/r_t=0.20$, infant weight-loss is higher than predicted for a tidal-field-free environment.  Note that the respective extents of these 3 regions depend on the tidal field strength (compare left and right panels).  Cluster-forming region mass-radius diagrams therefore constitute an exquisite tool to assess, in a glimpse, whether an external tidal field affects star cluster early evolution. 

If the CFRg mass-radius relation were one of constant mean surface density, the tidal field impact would increase with the mass of CFRgs by virtue of their decreasing mean {\it volume} density.  In other words, high-mass CFRgs are more prone to tidal overflow than their low-mass counterparts and $F_{bound}$ decreases with $m_{CFRg}$, thereby distorting the shape of the cluster mass function during violent relaxation \citep[see fig.~4 in][]{2010IAUS..266...87P}.  The most robust way of reproducing mass-independent tidal field impact $r_h/r_t$ is through constant mean {\it volume} density CFRgs (or, at least, mass-independent $\rho_{CFRg}$).  This is conducive to mass-independent infant weight-loss $1-F_{bound}$, and is in line with the observational constraint set by the invariant shape of the young cluster mass function.  {\it We therefore conclude that the mean volume density of cluster-forming regions is constant}.   \\

It may be worth stressing that the concepts `constant {\it mean} volume density' ($r_{CFRg} \propto m_{CFRg}^{1/3}$) or `constant {\it mean} {surface} density' ($r_{CFRg} \propto m_{CFRg}^{1/2}$), discussed in this contribution, are not to be misinterpreted as uniform (i.e. radially not varying) volume or surface density.  On the contrary, structures in molecular clouds are nested \citep{2010ApJ...712.1137K}, and molecular clumps have density gradients.
This property will be the starting point of the two topics discussed in Section~\ref{sec:clump}.    \\

The above conclusion that the mass-radius relation of CFRgs must be one of constant mean {\it volume} density stems from a purely dynamical line-of-reasoning.  This finding is independently confirmed by the tight linear correlation observed between the star formation rate and the {\it dense} molecular gas mass on galaxy scales \citep{2004ApJ...606..271G}, molecular-cloud scales \citep{2010ApJ...724..687L}, and molecular-clump scales \citep{2005ApJ...635L.173W}.  Here, `dense' means  hydrogen molecule number densities $n_{\rm H2} \simeq 10^{4-5}\,cm^{-3}$. 
This implies the existence of a number density threshold for star formation, $n_{{\rm H_2},th} \simeq 10^4\,cm^{-3}$.  It also implies that the mean number density of CFRgs is a few times $n_{{\rm H_2},th}$, the exact factor depending on the clump density profile \citep[see eq.~5 in][]{2011MNRAS.413.1899P}.  Therefore, the observational mapping of star formation in dense molecular gas also leads to $r_{CFRg} \propto m_{CFRg}^{1/3}$ \citep[see section 2 in][for a brief summary]{2011MNRAS.413.1899P}. \\

\section{Molecular Clumps versus CFRgs: Beware! \label{sec:clump}}

Observational mass-radius data sets of molecular clumps are to be handled with caution as they may be the imprint of the molecular tracer used to map them, rather than reflect cluster-formation conditions \citep[see section 3 in][for a discussion and enlightening examples]{2011MNRAS.411.1258P}.  For instance, figs.~9 and 21 in \citet{2010ApJS..188..313W} show the star-forming region W43S mapped in HCN 3$-$2 and CS 7$-$6 transitions, respectively.  While the FWHM and its enclosed mass are lower in CS 7$-$6 than in HCN 3$-$2, the corresponding mean volume density is about 5 times higher.  This is because CS 7$-$6 requires a higher volume density than HCN 3$-$2 to be excited.  Since molecular clumps have density gradients (see below), the CS 7$-$6 transition probes deeper inner regions of molecular clumps than HCN 3$-$2, hence the smaller mass and FWHM size inferred in CS 7$-$6 than in HCN 3$-$2.   
Besides, molecular line emissions extend smoothly beyond the FWHM size down to the noise level.  In other words, molecular clump FWHM sizes do {\it not} represent sharp physical boundaries \citep[see][and their fig.~25]{2010ApJS..188..313W}.  This illustrates that radii and masses of molecular clumps should {\it not} be interpreted as those of CFRgs.  This is the very reason for the specific and clear-cut terminology "cluster-forming region" used throughout this contribution.  

Molecular clumps have volume density profiles \citep[e.g.][]{2002ApJS..143..469M}:
\begin{equation}
\label{eq:profile}
\rho_{radial}(s) \propto s^{-p}\,,
\end{equation}
where $\rho_{radial}(s)$ is the volume density at the distance $s$ from the clump centre, and $p$ is the density index.

With star formation driven by a volume density threshold $n_{{\rm H_2}, th} \simeq 10^4\,cm^{-3}$, one can distinguish two zones in a molecular clump: a central CFRg, actively forming stars by virtue of a local number density higher than $n_{{\rm H2}, th}$, and an outer envelope inert in terms of star formation (see left panel of Fig.~\ref{fig:msf})\footnote{Another possible situation is that molecular clumps represent only a fraction of a CFRg because of a very high density molecular tracer}.  The mass of gas relevant to star formation is $m_{CFRg}$, rather than the overall clump mass $m_{clump}$, and the volume of star-forming gas does not coincide with the observed molecular clump.  This concept constitutes the crux of the two topics developed below.  

\subsection{Consequence 1 - Mass Functions: from Clumps to CFRgs \label{ssec:mf}}   

\begin{figure}[!t]
\includegraphics[width=15cm,angle=0,clip=true]{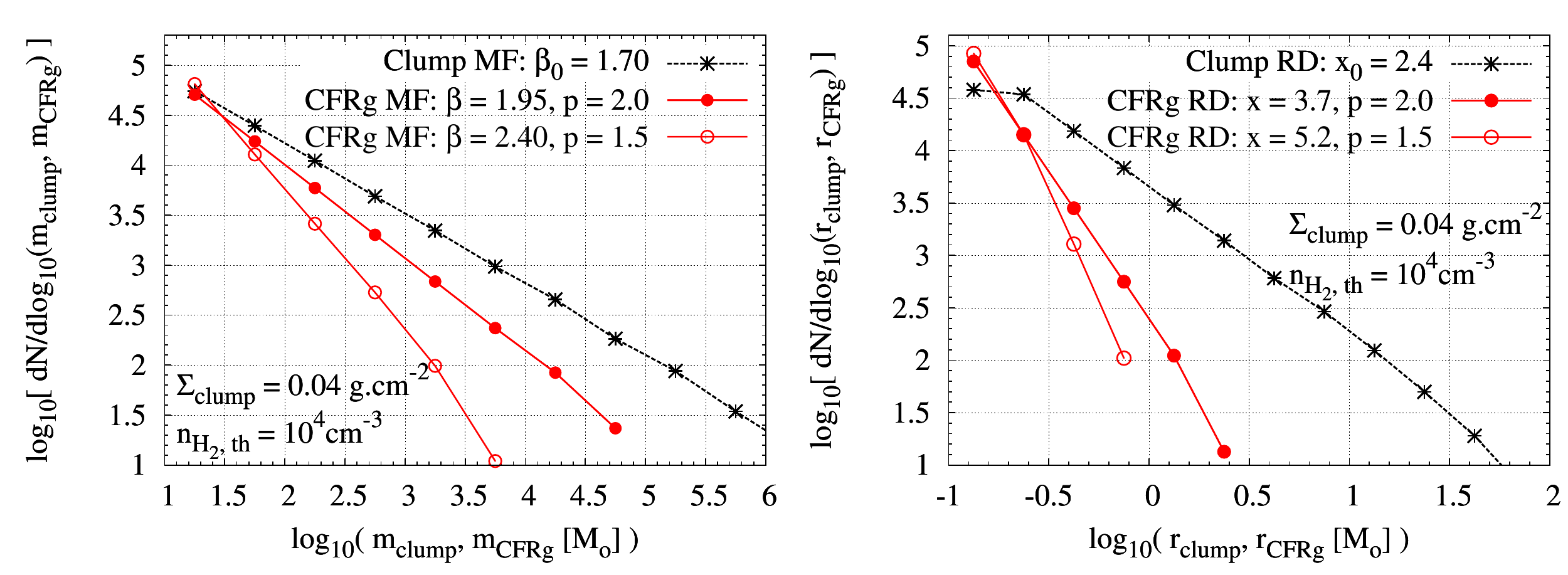} 
\caption{\label{fig:mf} Explaining why the mass functions (MF, {\it left panel}) and radius distributions (RD, {\it right panel}) of star clusters are steeper than those of molecular clumps and clouds.  In the model of relevance here, cluster-forming regions (CFRg) occupy a limited volume only of their host molecular clumps and are defined based on a number density threshold for star formation, $n_{H2,th}$ (see Fig.~\ref{fig:msf}, left panel).  The mean {\it volume} density of CFRgs is therefore constant.  If, in contrast, the clump mean {\it surface} density is constant, 
the difference in the mass-radius relations of clumps ($r_{clump} \propto m_{clump}^{1/2}$) and CFRgs ($r_{CFRg} \propto m_{CFRg}^{1/3}$) steepens the MF and RD of CFRgs compared to those of their host-clumps.  The degree of steepening depends on the clump density index $p$.  Note that $\Sigma_{clump}$ is the clump {\it mean} surface density. 
}
\end{figure}

It is puzzling that the young cluster mass function, $dN \propto m_\star^{-\beta_\star} dm_\star$, is steeper than the molecular cloud and clump mass functions, $dN \propto m_{clump}^{-\beta_0} dm_{clump}$, as  their respective indices are $\beta_{\star} \simeq 2$ and $\beta_0 \simeq 1.7$.  Since infant weight-loss is mass-independent, the mass function of young star clusters is also the mass function of embedded clusters, $dN \propto m_{ecl}^{-\beta_{ecl}} dm_{ecl}$ with $\beta_{ecl} = \beta_\star$.  
$\beta_0 \simeq 1.7$ and $\beta_{ecl} \simeq 2$ suggest that star formation is {\it less} efficient in high-mass clumps than in low-mass ones since the $SFE$ averaged over a clump, $SFE_{global}$, obeys: $SFE_{global} = m_{ecl}/m_{clump} \propto m_{clump}^{-0.3}$.  The slope $-0.3$ derives from $(\beta_0 - \beta_{ecl})/(\beta_{ecl} -1)$.  Accordingly, $SFE_{global}$ varies by {\it an order of magnitude} over 3 decades in clump mass, a lower limit to the cluster mass range in star cluster systems.  With the bound fraction $F_{bound}$ a very sensitive function of the $SFE$ \citep[][their fig.~1]{2007MNRAS.380.1589B}, a mass-varying $SFE$ does not seem a viable solution.
Assuming instantaneous gas expulsion and weak tidal field impact, $F_{bound}$ can vary virtually from 0 to almost unity when the $SFE$ varies by an order of magnitude \citep[fig.~1 in ][]{2007MNRAS.377..352P}.  Such 0-to-1 variations in the bound fraction of stars at the end of cluster violent relaxation are necessarily conducive to a severe reshaping of the cluster mass function through the first 100\,Myr of dynamical evolution \citep[see fig.~2 in][]{2008ApJ...678..347P}.  This is in stark disagreement with observed young cluster mass functions.  It would therefore be highly misleading to assume that the difference in slope between the mass functions of molecular clumps (or clouds) and young clusters is small enough ($\beta_{\star} - \beta_0 \simeq 0.3$) to lead to almost mass-independent $SFE$ and infant weight-loss.     

\begin{figure}
\includegraphics[width=15cm,angle=0,clip=true]{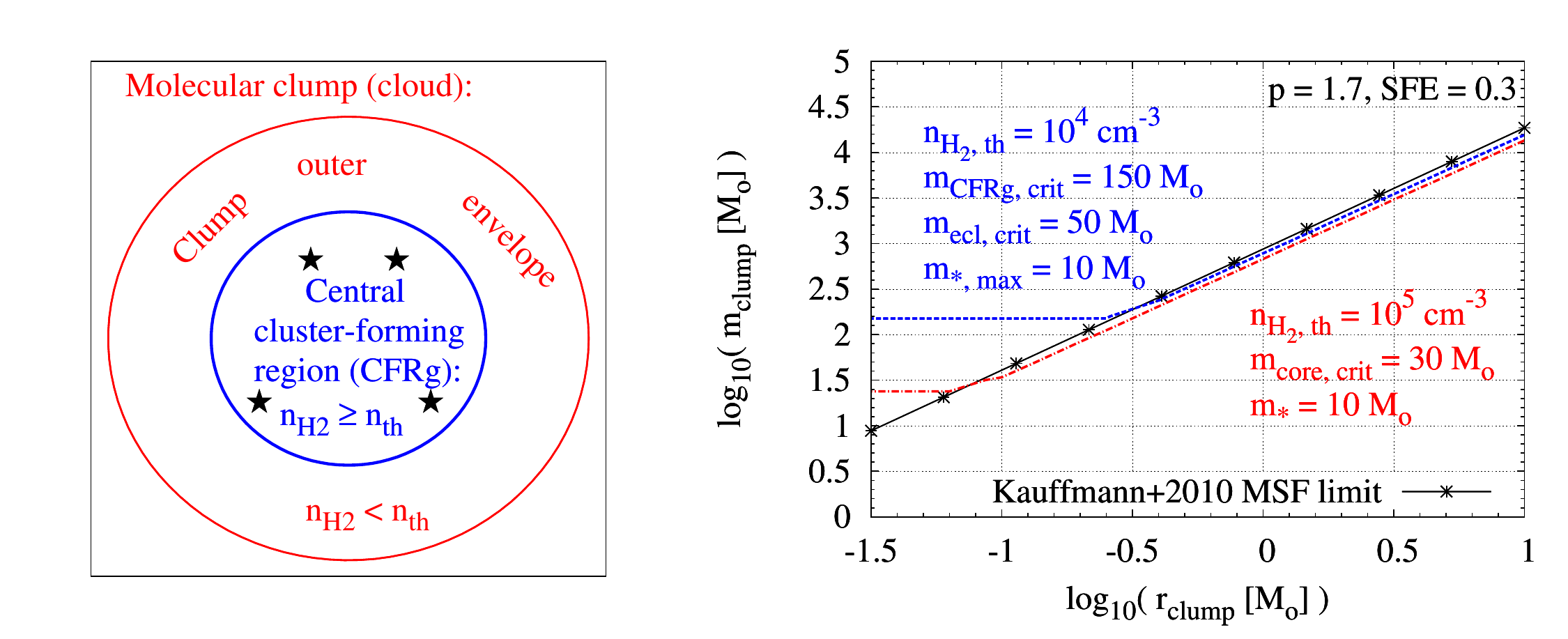} 
\caption{\label{fig:msf} {\it Left panel:} Defining a cluster-forming region (CFRg) nested {\it within} a molecular clump.  By virtue of the clump density gradient, the CFRg outer bound is defined by a volume (or number) density threshold for star formation, $\rho_{th}$ (or $n_{H_2, th}$).  {\it Right panel:} The observed massive star formation (MSF) limit (black solid line with asterisks).  It  equates with the mass-radius relation of molecular clumps containing the mass of star-forming gas (i.e.~$n_{H_2} \geq n_{H_2, th}$) needed to form a 10\,${\rm M}_{\odot}$-star, either on the small spatial scale ($r_{clump} < 0.2$\,pc) of an individual pre-stellar core (red dash-dotted line and lower-right data set), or on the larger scale ($r_{clump} \geq 0.2$\,pc) of a CFRg (blue dashed line and upper-left data set).   
}
\end{figure}

The issue we have to fix now is: how to reconcile mass-independent infant weight-loss with the difference in slope between the molecular clump and young star cluster mass functions.
The line of reasoning detailed above hinges on the identification of molecular clumps as CFRgs -- namely CFRg boundaries are those of molecular clumps.  In contrast, if CFRgs constitute a fraction only of the volume of their host molecular clump (left panel of Fig.~\ref{fig:msf}), star formation can be quantified by two distinct efficiencies {\it of different physical significances}.  The {\it global} SFE, defined on a clump-scale, is relevant to understand the difference between the cloud (clump) and young cluster mass functions.  We can also define an $SFE$ on the CFRg-scale.  This {\it local} SFE quantifies the ratio between the embedded-cluster stellar mass and the initial gas mass of the CFRg, $m_{ecl}/m_{CFRg}$.  This is the local $SFE$ -- {\it not} the global one -- which drives cluster violent relaxation.  The early dynamical evolution of star clusters and the mass function slope difference $\beta_{\star} - \beta_0$ are now dissociated issues.  Mass-independent infant weight-loss demonstrates that the local $SFE$ at the onset of gas expulsion is CFRg-mass-{\it in}dependent.  Consequently, the power-law mass functions of young star clusters, embedded-clusters and CFRgs all have the same slopes (i.e. $\beta_\star = \beta_{ecl} = \beta_{CFRg}$), by virtue of mass-independent infant weight-loss and mass-independent local $SFE$, respectively.  The mass function slope difference $\beta_{\star} - \beta_0$ tells us that the mass fraction of {\it dense star-forming gas} in molecular clumps, $m_{CFRg}/m_{clump}$, decreases with clump mass \citep[see fig.~1 in][for a summary-plot]{2011MNRAS.413.1899P}.  What can be the reason for such a behaviour?

The mean surface density of GMCs (the precursors of massive star clusters forming profusely in galaxy mergers) is observed to be about constant, in our Galaxy and in the Magellanic Clouds \citep[see fig.~8 in][]{2007prpl.conf...81B}.  This property may also characterize {\it star-forming} molecular clumps \citep[see fig.~10 in][]{2011MNRAS.413.1899P}, where Galactic disc star clusters form.  
It therefore appears that the mass-radius relation of CFRgs on the one hand, and the mass-radius relations of molecular clumps and GMCs on the other, have different slopes with: $r_{CFRg} \propto m_{CFRg}^{1/3}$, while $r_{clump} \propto m_{clump}^{1/2}$.  {\it This immediately implies a mass-dependent effect upon the ratio $m_{CFRg}/m_{clump}$}.  As we already saw from Fig.~\ref{fig:tf}, a constant mean surface density is conducive to a mean volume density decreasing towards higher masses.  As a result, the clump mass fraction of gas denser than a given number density threshold is also a decreasing function of the clump mass.  Considering the number density threshold for star formation, $n_{{\rm H2}, th}$, the desired trend follows logically: $m_{CFRg}/m_{clump}$ decreases with $m_{clump}$, which steepens the CFRg mass function compared to the clump mass function.  

The degree of steepening depends on the clump density index $p$: the shallower the clump density profile, the greater the difference in slope $\beta_{CFRg} - \beta_0$.  The left panel of Fig.~\ref{fig:mf} illustrates this effect for $p=2$ (isothermal sphere) and $p=1.5$.  The (black) dashed line with asterisks depicts the clump mass function, the (red) solid lines the resulting CFRg mass functions (full and open symbols: $p=2.0$ and $p=1.5$).  The mean volume density of CFRgs is $\rho_{CFRg} = [3/(3-p)] \times \rho_{th}$, where $\rho_{th}$ is the volume density threshold for star formation \citep[eq.~5 in][]{2011MNRAS.413.1899P}.  The mean surface density of their host molecular clumps is assumed to be constant $\Sigma_{clump}=0.04\,g.cm^{-2}$.  Figure~6 in \citet{2011MNRAS.413.1899P} shows that to steepen the clump mass function $\beta_0 \simeq 1.7$ into that of CFRgs (hence of embedded clusters), $\beta_{CFRg} \simeq 2$, a density index $p \simeq 1.9$ is required.  This is in excellent agreement with the mean density index inferred via dust-continuum mapping of star-forming regions by  \citet{2002ApJS..143..469M}.  Therefore, the model does a great job at explaining the observed slope difference $\beta_{\star} - \beta_0$.  {\it We conclude that the difference in slope between the molecular clump and young star cluster mass functions can arise from different mass-radius relations for CFRgs and their host molecular clumps.}

Not only does this effect steepens the mass function, it also steepens the {\it radius} distribution, as illustrated in the right panel of Fig.~\ref{fig:mf} (same model parameters and same colour/symbol-codings as in left panel).  $x_0$ and $x$ are the spectral indices of the radius distributions of clumps and CFRgs, respectively: $dN \propto r_{clump}^{-x_0} dr_{clump}$ and $dN \propto r_{CFRg}^{-x} dr_{CFRg}$.
This model has therefore the potential to explain why the observed GMC mass function {\it and} radius distribution are shallower than their young star cluster counterparts \cite[see section 5 in][for a full discussion of the radius distribution aspect]{2011MNRAS.413.1899P}. 

\subsection{Consequence 2 - The Massive Star Formation limit \label{ssec:msf}}

\citet[][top and middle panels of their fig.~2]{2010ApJ...723L...7K} recently highlighted the presence of a limit for massive star formation (MSF) in the mass-size space of molecular structures.  This MSF limit obeys: 
\begin{equation}
\label{eq:kaulim}
m(r)=870M_{\odot}(r/pc)^{1.3}\,,
\end{equation}
where $m(r)$ is the gas mass enclosed within the effective radius, $r$. 
It is shown in the right panel of Fig.~\ref{fig:msf} as the (black) solid line with asterisks.  Molecular structures lying below the MSF limit are void of massive stars, while those more massive than the MSF limit show HII regions and, therefore, contain stars more massive than 8-10\,${\rm M}_{\odot}$.  
In other words, MSF demands a mass of molecular gas enclosed within any given projected radius higher than what Eq.~\ref{eq:kaulim} predicts.

Armed with the model presented in Section~\ref{ssec:mf}, \citet{2011MNRAS.tmp..939P} present an original solution for the origin of the MSF limit.  The key-idea is to successively relate: \\
{\it (1)}~the mass of a molecular clump, $m_{clump}$, to the mass of the CFRg, $m_{CFRg}$, it contains, \\
{\it (2)}~the CFRg mass to the final mass of the embedded cluster it forms, $m_{ecl}$, and \\
{\it (3)}~the mass of the embedded cluster $m_{ecl}$ to the most probable mass of the most-massive star it contains, $m_{*, max}$.  Figure~3 of \citet{2010MNRAS.401..275W} illustrates the observed relation between the mass  of young or embedded clusters, and the mass of their most massive star.  As such, it 
provides us with a solution to Step~3: the formation of a star of mass $m_{*,max} \geq 8-10\,{\rm M}_{\odot}$ requires $m_{ecl} \geq 50\,{\rm M}_{\odot}$.  Step~2 simply equates with the definition of the local $SFE$, that is, $m_{ecl} = SFE \times m_{CFRg}$, as seen in Section~\ref{ssec:mf}.  With the `canonical' value, $SFE \simeq 0.3$, this gives a limit $m_{CFRg, crit} \simeq 150M_{\odot}$ of dense star-forming gas for the formation of a $10\,{\rm M}_{\odot}$-star.  Step~1 is implemented through eq.~3 of \citet{2011MNRAS.413.1899P}, reproduced below for the sake of clarity.  It relates the mass $m_{clump}$ and radius $r_{clump}$ of molecular clumps to the CFRg mass they contain\footnote{Note that in \citet{2011MNRAS.413.1899P}, the CFRg mass and radius are referred as $m_{th}$ and $r_{th}$, respectively}: 
\begin{equation}
\label{eq:mcfrg}
m_{CFRg} = \left( \frac{3-p}{4 \pi \rho _{th}} \right)^{(3-p)/p} m_{clump}^{3/p} \, r_{clump}^{-3(3-p)/p}\,,
\end{equation}
or: 
\begin{equation}
\label{eq:mr}
m_{clump}=m_{CFRg}^{p/3} \left( \frac{4 \pi \rho_{th} }{3-p} \right)^{(3-p)/3} r_{clump}^{3-p}\;.
\end{equation}
For given density index $p$ and star-formation density threshold $\rho_{th}$, Eq.~\ref{eq:mr} defines the mass-radius relation of molecular clumps containing a given CFRg mass, $m_{CFRg}$.  [This mass-radius relation is of the same nature as the iso-$m_{th}$ lines shown in top panel of fig.~3 in \citet{2011MNRAS.413.1899P} where $r_{clump}$ is plotted against $m_{clump}$ and $m_{th} \equiv m_{CFRg}$].

\begin{table}[t] 
\caption{Density index $p$ of molecular clumps (see Eq.~\ref{eq:profile})} 
\center
\begin{minipage}{0.97\textwidth}
\center
\begin{tabular}{cll} 
\hline\hline 
$p$ & Method & Source \\ [0.5ex] \hline   
$p=1.9$ & Mass function steepening      & Section~\ref{ssec:mf}\\ 
$p=1.7$ & Massive star formation  limit & Section~\ref{ssec:msf}\\
$<p>=1.8\pm0.4$ & Star-forming clump dust-continuum mapping        & \citet{2002ApJS..143..469M} \\ [1ex]  
\hline
\end{tabular} 
\end{minipage}
\label{tab:p} 
\end{table}

We can therefore identify the MSF limit, Eq.~\ref{eq:kaulim}, to Eq.~\ref{eq:mr} with $m_{CFRg, crit} \simeq 150\,{\rm M}_{\odot}$.  This gives $p=1.7$ and $n_{H_2, th} \simeq 10^4\,cm^{-3}$.  The inferred number density threshold is in remarkable agreement with values previously suggested in the literature (see end of Section~\ref{sec:mrr}).  The density index $p$  inferred from the MSF limit is also very similar to observed density indices of molecular clumps, and to the $p$-value needed to steepen the clump mass function index $\beta_0$ into that of young star clusters $\beta_\star$ (see Section~\ref{ssec:mf} and Table~\ref{tab:p}). 
{\it We therefore conclude that the observationally inferred MSF limit equates with the mass-radius relation of molecular clumps containing the mass of dense star-forming gas needed for the formation of a $8-10\,{\rm M}_{\odot}$ star}.
Figure~\ref{fig:msf} shows Eq.~\ref{eq:mr} with $m_{CFRg} = 150\,{\rm M}_{\odot}$, $p=1.7$ and $\rho_{th} = 700\,{\rm M}_{\odot}$ ($\equiv n_{H_2, th} \simeq 10^4\,cm^{-3}$) as the (blue) dashed line.  Note that the result depends weakly only on the assumed local $SFE$ since $m_{clump} \propto m_{CFRg}^{p/3} \propto (m_{ecl}/SFE)^{p/3}$ in Eq.~\ref{eq:mr}.  \\

The formation of massive stars as star cluster members allows us to explain the observed MSF limit down to a spatial scale of $\simeq 0.2$\,pc.  Below that limit, the entire clump gas is denser than the threshold $n_{H_2, th} \simeq 10^4\,cm^{-3}$ which causes the model to depart from the observed MSF limit.  Over the range $< 0.2$\,pc, a similar line of reasoning shows that the MSF limit is consistent with the mass a pre-stellar core\footnote{We adhere to the following nomenclature: the word `core' refers to the gaseous precursor of an individual star or of a small group of stars, while the term `clump' is designated for regions hosting cluster formation.} must have to form a $10\,{\rm M}_{\odot}$ star.  This `individual-star-formation' picture is actually expected if observations look {\it into} a forming-star-cluster, at the spatial scale of individual pre-stellar cores.  Pre-stellar cores correspond to density peaks of at least $n_{{\rm H_2}, th}=10^5\,cm^{-3}$ because stars form fastest where $n_{\rm H2} > 10^5\,cm^{-3}$ \citep[see][his section 3.6]{2007ApJ...668.1064E}.  The comparison between the core mass function and the stellar IMF performed by \citet{2007A&A...462L..17A} for the Pipe dark cloud gives $SFE_{core} \simeq 0.3$.
The formation of a $10\,{\rm M}_{\odot}$-star thus requires a mass of star-forming gas $m_{core, crit} = 30\,{\rm M}_{\odot}$.  The (red) dash-dotted line in the right panel of Fig.~\ref{fig:msf} depicts Eq.~\ref{eq:mr} with $m_{core, crit} = 30\,{\rm M}_{\odot}$, $p=1.7$ and $\rho_{th} = 7000\,{\rm M}_{\odot}$ ($\equiv n_{H_2, th} \simeq 10^5\,cm^{-3}$).  Obviously, the observed MSF limit is also consistent with the formation of a $10\,{\rm M}_{\odot}$-star out of its individual density peak with $n_{th} \simeq 10^5\,cm^{-3}$. 
Note that the model does not allow to disentangle between individual star formation in the field, or individual star formation in clusters.  
{\it The observed MSF limit therefore embodies information about the formation of massive stars as star cluster members, and the formation of massive stars out of their individual pre-stellar cores.  In this framework, the density threshold $n_{{\rm H_2}, th} \simeq 10^4\,cm^{-3}$ for clustered star formation probably represents the mean density above which the formation of local density peaks with $n_{{\rm H_2},th} \simeq 10^5\,cm^{-3}$ is favoured in supersonically turbulent gas} \citep[see][]{2011ApJ...731...61E}.    \\

Note that the slope of the MSF limit in the $r_{clump}-m_{clump}$ space is 1.3.  Neither is the MSF limit a relation of constant mean surface density, nor is it a relation of constant mean volume density (although the model hinges on a volume density threshold for star formation).  The one property common to all clumps along the MSF limit is the mass of star-forming gas, either on a star-cluster-scale ($m_{CFRg} \simeq 150M_{\odot}$, $n_{th} \simeq 10^4\,cm^{-3}$), or on a pre-stellar core scale  ($m_{core} \simeq 30M_{\odot}$, $n_{H_2, th} \simeq 10^5\,cm^{-3}$).

\section{Conclusions\label{sec:conclu}}

The analysis of the impact exerted by a tidal field upon star clusters which have expelled their residual star-forming gas has been carried out.  It shows that the time-invariant shape of the young star cluster mass function is robustly reproduced if the mean {\it volume} density of cluster-forming regions (CFRgs) is constant.  If the mass-radius relation of CFRgs were one of constant mean {\it surface} density, it would lead to the preferential removal of high mass-clusters, which contradicts observational results.  These trends follow from how CFRg volume density and CFRg mass scale with each other.  This stellar-dynamics-based  finding is independently confirmed by studies mapping star formation activity in molecular clumps, molecular clouds and galaxies, which all show that star formation is driven by a volume (number) density threshold.

That star formation takes place in gas denser than a given number density threshold, $n_{{\rm H_2},th}$, and that molecular clumps are characterized by density gradients allow us to define two distinct regions in molecular clumps: a central dense CFRg, and an outer envelope inert in terms of star formation due to $n_{{\rm H_2}}<n_{{\rm H_2},th}$.  As a result, properties (e.g. mass, radius, mean volume densities) of CFRgs and molecular clumps get dissociated. 

Building on that picture, I put forward an original explanation for why the mass function of young star clusters is steeper than that of molecular clumps and clouds.  The mass-radius relation of molecular clouds and, possibly, of star-forming clumps, is one of constant mean surface density.  This contrasts with the constant mean volume density inferred for CFRgs.  This difference in slope in their respective mass-radius relations steepens the mass function of CFRgs compared to that of their host-clumps because clumps of higher mass have a lower mean volume density (hence a smaller fraction of star-forming gas) at constant mean surface density.

Finally, the same model is successfully applied to understand the origin of the massive star formation limit (MSF) in the mass-size space of molecular structures.  
This is shown to be consistent with a {\it threshold in star-forming gas mass} beyond which the star-forming gas reservoir is large enough to allow the formation of massive stars.  Specifically, the MSF limit is consistent with the formation of a $10\,{\rm M}_{\odot}$-star out of its individual density peak with $n_{{\rm H_2}, th} \simeq 10^5\,cm^{-3}$, or with the formation of a $10\,{\rm M}_{\odot}$-star as a CFRg member with $n_{{\rm H_2}, th} \simeq 10^4\,cm^{-3}$. \\

Slides of oral presentations related to the topics discussed in this contribution are available at
\href{http://www.astro.uni-bonn.de/~gparm/talks.html}{http://www.astro.uni-bonn.de/$\sim$gparm/talks.html}

%
%
\small  
%
\section*{Acknowledgments}   
This work has been supported by a Research Fellowship of the Humboldt Foundation and a Research Fellowship of the Max-Planck-Institut f\"ur Radioastronomie, Bonn, Germany.
%
\bibliographystyle{aa}

%


\end{document}